\documentclass[aps,prl,twocolumn,showpacs]{revtex4}
\usepackage{amssymb}
\usepackage{amsfonts}
\usepackage{amsmath}
\usepackage{graphicx}
\usepackage{CJK}
\begin{document}
%\begin{CJK}{GBK}{song}

%draft

\title{Topological Phase Transition and Chiral-Anomaly Driven Negative Magneto-Resistance in Bulk Black Phosphorus}

\author{Chun-Hong Li$^1$, Yu-Jia Long$^1$, Ling-Xiao Zhao$^1$, Lei Shan$^{1,4}$, Zhi-An Ren$^1$, Jian-Zhou Zhao$^{1,3}$, Hong-Ming Weng$^{1,4}$, Xi Dai$^{1,4}$, Zhong Fang$^{1,4}$, Gen-Fu Chen$^{1,4,\dag}$, Cong Ren$^{1,2,\ddag}$}

\affiliation{$^1$ Beijing National Laboratory for Condensed Matter Physics, Institute of Physics, Chinese Academy of Sciences, P.O. Box 603, Beijing 100190, China}

\affiliation{$^2$ Physics Department, Yunnan University, Kunming 671000, China}

\affiliation{$^3$ Co-Innovation Center for New Energetic Materials, Southwest University of Science and Technology, Mianyang, Sicuan 621010, China}

\affiliation{$^4$ Collaborative Innovation Center of Advanced Microstructures, Beijing 100190, China}

\begin{abstract}
We report the anisotropic magneto-transport measurement on a non-compound band semiconductor black phosphorus (BP) with magnetic field \textit{\textbf{B}} up to 16 Tesla applied in both perpendicular and parallel
to electric current \textit{\textbf{I}} under hydrostatic pressures.  The BP undergoes a topological Lifshitz transition from band semiconductor to a zero-gap Dirac semimetal state, characterized by a weak localization-weak antilocaliation transition at low magnetic fields and the emergence of a nontrivial Berry Phase of $\pi$ detected by SdH magneto-oscillations in magnetoresistance curves.  In the transition region, we observe a pressure-dependent negative MR only in the \textit{\textbf{B}}$\parallel$\textit{\textbf{I}} configuration.  This negative longitudinal MR is attributed to the Adler-Bell-Jackiw anomaly (topological \textit{\textbf{E}}$\cdot$ \textit{\textbf{B}} term) in the presence of weak antilocalization corrections.
\end{abstract}

\pacs{74.20.Rp, 74.25.Ha, 74.70.Dd}

\maketitle

\newpage

More recently a new kind of topological materials termed Dirac or Weyl semimetal, three-dimensional (3D) analogs of two-dimensional graphene, has been intensively investigated both theoretically and experimentally in that Dirac or Weyl semimetal is a phase of matter that provides a solid state realization of chiral Weyl fermions
\cite{Chirality1,Chirality2,Chirality2a,3Diracmetal1,3Da,3Dirac,3Diracmetal2,3Diracmetal2a}.
Most of its unique physics is a consequence of chiral anomaly, namely non-conservation of the number of quasiparticles of a given chirality.  This extraordinary property is notably characterised by a large and strongly anisotropic negative magneto-resistance (MR) which exists in the case when the electric and magnetic fields are
collinearly aligned.  Indeed, following the theoretical prediction, the chiral anomaly-induced negative MR has been notably realized in Dirac semimetal Cd$_3$As$_2$ \cite{3Diracmetal3,Ong1} and Na$_3$Bi \cite{Ong2}, Weyl semimetal TaAs \cite{TaAs1,TaAs2}, ZeTe$_5$ \cite{ZrTe5} and noncentrosymmetric Weyl semimetals NbAs
\cite{NbAs} and NbP \cite{NbP}. However, a sizable negative MR also exist in semimetals which lack Dirac linear dispersion, such as Cd$_x$Hg$_{1-x}$Te \cite{CdHgTe}, PdCoO \cite{PdCoO} and half-Heusler GdPtBi \cite{GdPtBi}.  The question naturally arises whether the chiral-related negative MR is an intrinsic property of Dirac or Weyl semimetal. Therefore, it is desirable to search for the origin of negative MR in a broader class
of semimetals.

Recently the narrow band-gap semiconductor, black phosphorus (BP), has been revived owing to the realization of mono-layered crystalline structure (phosphorene) and the exhibition of promising carrier mobilities, possible a new candidate for next-generation electronic and spintronic devices \cite{BPuse1,BPuse1a,BPuse2,BPuse2a,BPuse3}. Fundamentally, BP has a relatively low band gap which can be further reduced by increasing the interlayer coupling. As its counterparts, slight change in the crystal structure thus strongly modifies the band gap of BP
\cite{Morita,Narita}. In particular, first-principle calculation predicts that BP posses a unique band structure, whose dispersion is nearly linear along the armchair direction \cite{Ptheory1,Ptheory2}. Recent photoemission and magneto-transport measurements appear to support the theoretical prediction that bulk BP host 3D Dirac
semimetal phase \cite{BPexperim1,BPexperim}.  Therefore, in such Dirac semimetal of BP, there is strong interest in whether the chiral anomaly can be detected as a negative contribution to the longitudinal MR.  In this paper we apply a moderate hydrostatic pressure to drive bulk BP into a semimetallic state. By anisotropic magnetoresistance
measurements we observe a large negative MR only in the presence of electric and magnetic fields aligned collinearly.  This negative longitudinal MR is attributed to the Adler-Bell-Jackiw anomaly (topological \textit{\textbf{E}}$\cdot$ \textit{\textbf{B}} term) in the presence of weak antilocalization corrections.

\begin{figure}
\includegraphics[scale=0.4]{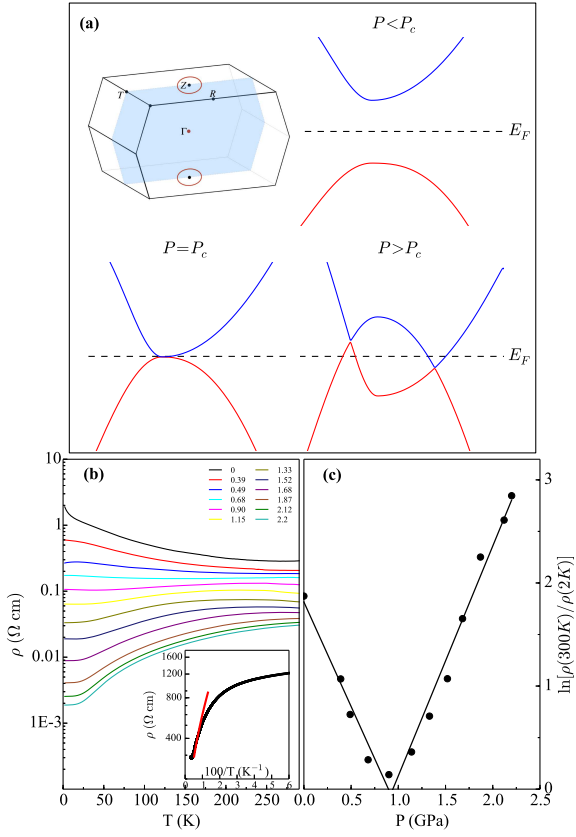}
\caption{\label{fig:fig1}(Color online) (a) Brillouin zone and the evolution of the band structure diagram of the black phosphorus under hydrostatic pressures.  Band structures are calculated by using the Vienna ab initio simulation package (VASP) based on generalized gradient approximation in Perdew-Burke-Ernzerhof (PBE) pseudo-potentional, the spin-orbit coupling (SOC) is included in our calculation. (b) Transverse resistivity
$\rho_{xx}$ as a function of temperature $T$ under several
hydrostatic pressures. Inset: The semi-logarithmic plot of
resistivity vs. $1/T$ at ambient pressure.  The red solid line is
the linearly fitting line. (c) The semi-logarithmic plot of
residue-resistance-ratio [$RRR\equiv R(300K)/R(2K)$] as a function
of $P$.  The solid lines are linear fits in the limited pressure
region, respectively.}
\end{figure}

\begin{figure}
\includegraphics[scale=0.35]{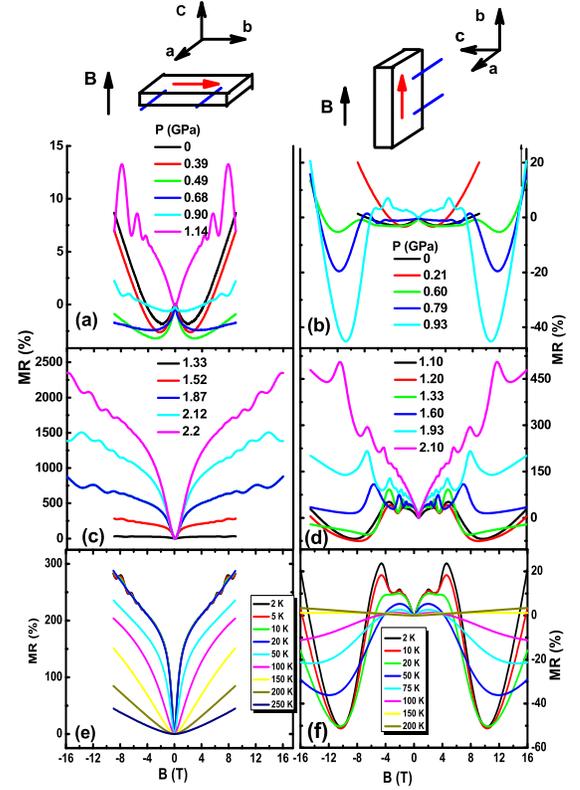}
\caption{\label{fig:fig1}(Color online)(a) (c) Pressure dependence
of magnetoresistance ($MR\equiv \frac{\rho(B)-\rho(0))}{\rho(0)}$)
at $T=2$ K with magnetic field $B$ parallel to $c$-axis. (b) (d)
Pressure dependence of magnetoresistance at $T=2$ K with magnetic
field $B$ parallel to $b$-axis. The upper insets of (a) and (b) are
schematic of the measurement configuration for $B\parallel c$ and $B\parallel b$,
respectively. The red arrowed lines show the electric current with
respect to the applied magnetic field. (e) Temperature dependence of
MR with $B\parallel c$ under $P=1.52$ GPa. (f) Temperature dependence of MR
with $B\parallel b$ under $P=1.20$ GPa.}
\end{figure}

Fig. 1(a) schematically illustrates the pressure dependence of band structures of bulk BP using first-principle electronic structure calculations.  As calculated, the semiconducting system opens a gap of several hundred milli-eV  at ambient pressure.  With increasing pressure $P$, the semiconducting band gap gradually closes and the two bands below and above Fermi energy $E_F$ touch at the $Z$-point.  The band inversion happens when further increasing $P$, and inversion gaps are reopened at crossing points due to finite spin-orbit coupling. Experimentally, the BP crystal is a $p$-type semiconductor with an activation energy of $E_g\simeq 8.6 $ meV in resistivity $\rho$ vs. temperature ($T$) at ambient pressure, as shown in the inset of Fig. 1(b). Upon the applied pressure $P$ the resistive divergence at low $T$s is gradually suppressed and a completely metallic state in the whole $T$ range, occurs as $P \geq 1.52$ GPa, as shown in the main penal of Fig. 1(b).  This dramatic resistivity change with $P$-driven semiconductor-to-metal (STM) transition was also observed in 3D topological Kondo insulator SmB$_6$ near a quantum critical pressure of 5.4 GPa \cite{SmB6}.  For a rough estimate of the critical pressure $P_c$ for STM transition in our BP samples, we choose residue-resistance-ratio [$RRR\equiv R(300K)/R(2K)$] as a function of $P$, as shown in Fig. 1(c).  In this scenario, the critical pressure for STM transition is between 0.8 and 1.1 GPa.  Moreover, as shown in Fig. 1(c), $RRR$ scales with $P$ as $RRR\propto \exp(-aP)$ on the semiconducting side ($P<P_c$) implying a $P$-dependent band gap and eventually, a zero-gap semiconductor at $P_c$ for BP. Similar result has been obtained in $P$-dependent $\rho-T$ measurement \cite{Morita}.

To identify the nature of the $P$-induced STM transition and to provide useful information on Fermi surface topology for metallic BP, we performed magnetoresistance measurements with magnetic field $B$ applied parallel ($B\parallel c$) and perpendicular ($B\parallel b$) to the basal $c$-plane, respectively. It is noted that in parallel $B\parallel c$ the magnetic field is perpendicular to the current $I$, transverse magnetoresistance TMR.  On the contrary, in perpendicular $B$ is in the parallel direction of $I$, namely, $B\parallel I$ the so-called Lorentz-force-free configuration, resulting in the longitudinal magnetoresistance LMR.  Fig. 2 summarizes the main results of magnetoresistance (MR) measurement at varied $P$s and $T$s with $B\parallel c$ and $B\parallel b$, respectively.  At $P\leq 0.9$ GPa in Fig. 2(a) and (b), the MR$\equiv\frac{\rho(B)-\rho(0)}{\rho(0)}$ data show a sharp peak (negative) MR at low fields for both $B\parallel c$ and $B\parallel b$. In this pressure region, the negative MR in the low field is attributed to weak localization (WL) effect in band semiconducting state \cite{PhysReport,WL-theory,WangYY,WangYYa}. However, as $P$ approaches to $P_c=1.14$ GPa, the negative MR completely vanishes, leaving a relatively sharp positive MR superimposed on a underlying magneto-oscillations for $P\geq P_c\simeq 1.14$ GPa, as clearly illustrated in Fig. 2(a). These dips in MR are similar to those observed in thin film of magnetically doped Bi$_2$Se$_3$ topological insulators \cite{WangYY,WangYYa,BiSb-ABJ}, graphene \cite{WAL-Graphene,WAL-Graphenea} and 3D Dirac semimetal Cd$_3$As$_2$ \cite{CrAs-WAL}, which is a characteristic feature of weak antilocalization (WAL) effect \cite{WAL-Graphenea}.

Generally, WL-WAL transition at low magnetic field is a consequence of the emergence of topological band structure.  The destructive interference due to the $\pi$ Berry phase in momentum $k$-space can give an enhancement to the classical electronic conductivity in small magnetic fields, leading to a peculiar weak antilocalization effect \cite{WAL-Graphene,WAL-Graphenea}.  In addition, the emergence of a nontrivial Berry phase at $P > P_c$ has been confirmed by analysing the magneto-oscillations imposing on the MR curves at high magnetic fields, see Supplement Material SII. Intriguingly, the WL-WAL crossover at low fields and the appearance of strong magneto-oscillations at high magnetic fields at such $P_c$ are accompanied with a dramatic enhancement of the MR ratio, reaching a large value of 2200\% in 16 T with no indication of saturation, as shown in Fig. 2(a). Similar large positive MR for $B\parallel c$ has been reported in a number of semimetals, such as WTe$_2$ \cite{WTe2}, TaAs \cite{TaAs1,TaAs2} and Bismuth \cite{Bismuth}, and has been attributed to the electron-hole compensation effect. Very recently, a titanic non-saturating MR of 80000\% in a more insulating BP sample has been observed and coincident with a sign reversal of Hall effect from negative (electron-type) to positive (hole-type) transition at $P_c\simeq 1.2$ GPa \cite{BPexperim}.  Combining with all of these experimental observations, including a $P$-dependent zero-band gap state, the WL-WAL crossover and the emergence of an electron pocket with nontrivial Berry phase point to a topological phase transition of the BP band structure at $P_c$.

\begin{figure}
\includegraphics[scale=0.40]{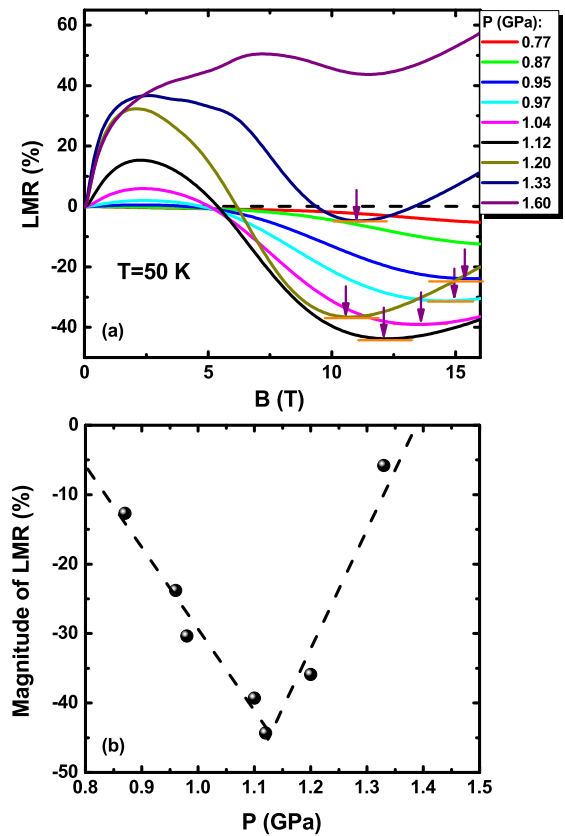}
\caption{\label{fig:fig2}(Color online)(a) Pressure dependence
of MR in the parallel configuration ($B\parallel b$). The arrows mark the trace of the turning points in such LMR curves (the minimum level of the LMR) (b) The magnitude of negative MR as a function of applied pressure at $T=50$ K. The dashed lines are linear fits in the limited pressure region, respectively.}
\end{figure}

In Fig. 2(b) and 2(d), a distinguishing feature is that in longitudinal MR curves with $B\parallel I$, a large  negative MR appears at relatively high fields between 4 T and 8 T in pressure region of $0.6\sim 1.6$ GPa. Under higher $P$, the negative LMR disappears even in Dirac semimetal state, restoring to a normally quadratic MR with Landau oscillation level.  It can be seen that the negative LMR is beside on the WAL, reaching a maximum value of -60\% before a upturn to positive LMR at higher magnetic fields at $P=1.2$ GPa.  To clearly observe the negative LMR, we increase $T$s to suppress the magneto-oscillations in MR. As shown in Fig. 2(f), below 20 K, the level of the negative LMR is nearly $T$-independent.  With increasing $T$, both the negative LMR and the WAL effect are suppressed, and ultimately disappear above 150 K. In contrast, shown in Fig. 2(c) and Fig. 2(e) the TMR curves show fully positive values in semimetal state ($P\geq P_c$), even no tendency toward a negative MR both in $P$s and in $T$s. To double check the occurrence of the negative MR only in the parallel configuration, we also measured the MR curves with $B\parallel a$-axis up to 16 T and the electric current always in $ab$-plane under same pressures (see Supplement Material SIII). In these transverse MR curves, no trace of negative MR is found.

From MR point of view, the appearance of negative LMR is rare for non-ferromagnetic systems due to the absence of Lorentz force acting on electron/quasiparticle system.  However, for topological materials it is believed that negative LMR is a signature of the chiral anomaly, i.e., nonconservation of the chiral charge in the present of  collinear-oriented \textbf{\textit{E}} and \textbf{\textit{B}}.  Burkov \cite{Chiraltheory2} argued that the occurrence of chiral anomaly-driven negative LMR has two crucial ingredients.  One is the magnetic field-induced coupling between the chiral and the total charge densities.  This arises from the Berry curvature and is present in principle whenever the Berry curvature is nonzero.  In this case the observation of negative LMR is nonspecific to Dirac and Weyl metals. However, only when the chiral charge density is a nearly conserved quantity, the coupling between the chiral and the total charge densities leads to a large negative LMR.  This property is specific to Dirac and Weyl metals and is realized only when the Fermi energy is close to Weyl nodes.

To demonstrate the chirality-related nature of the negative LMR of BP, we show in Fig. 3 (a) the LMR curves under various $P$s at a elevated $T=50$ K. As shown, the negative LMR is strongly $P$-dependent in magnitude.  At $P < P_c$, the level of the negative LMR increases with $P$, as arrow-marked in Fig. 3(a).  At $P_c=1.12$ GPa, the negative LMR reaches a maximum of -40\% in magnitude, and then it decreases and completely disappears, entering a positive LMR state at $P>1.33$ GPa. Fig. 3(b) shows the linear dependence of the magnitude of negative LMR on $P$.  This behavior of the negative LMR is similar to the topological phase transition based on RRR criterion [Fig. 1(c)], indicative of the touching/closing of the band gap reaching a generic Weyl node point at $E_F$ in band structure of 3D Dirac semimetals. This Weyl node-related negative LMR is the most prominent signature in magnetotransport for the chiral anomaly for BP.

For a quantitative estimate of the chiral anomaly-induced negative LMR, we attribute the underlying mechanism for the negative MR to the Adler-Bell-Jackiw (ABJ) anomaly in the presence of WAL corrections. Based on a semi-classical theory of motion for the momentum, Kim \textit{et al.} developed the equation describing the ABJ anomaly and possible scattering channels of motion for the momentum \cite{BiSb-ABJ,Chiraltheory1a}. In their model, the longitudinal magnetoconductivity (MC) in the weak field region is expressed as:
\begin{equation}
\sigma_L(B)=(1+C_WB^2)\cdot \sigma_{WAL} +\sigma_n,
\end{equation}
with $\sigma_{WAL}$ the conductivity from WAL corrections associated with scattering and $\sigma_n$ being that from conventional Fermi surface contributions.  Here the factor $C_{W}B^2$ with a positive constant $C_W$ originates from the topological $(\bf{E}\cdot \bf{B})\Omega_b$ term in the equation ($\Omega_b$ the Berry curvature in momentum).  It is noted that a $\sim B^2$ term of chiral magnetic effect for topological systems has been deduced theoretically by Son \textit{et al}. \cite{CME} and Kharzeev \textit{et al}. \cite{CME1}. On the other hand, the transverse MC is expressed as $\sigma_T(B)=\sigma_{WAL}+\sigma_n$ without the anomaly contribution because of vanishing of the contribution from $\textbf{E}\cdot \textbf{B}$ term.  Fig. 4(a) and 4(b) show the typical transverse MC and longitudinal MC curves at different $T$s in the limited magnetic field region of $-4 < B < 4$ T, respectively, with their correspondingly theoretical fitting based on the above equations. As shown, the theoretical fits to the data reproduce quite well the essential features of these MC curves, yielding the important parameters as: $C_W=1.39 \times 10^{-2}\ T^{-2}$ for $T=30$ K and $C_W=1.0 \times 10^{-2}\ \ T^{-2}$ for $T=50$ K for the longitudinal MC curves.  From these parameters, it can be sensed that with increasing $T$, the value of $C_W$ deceases. As expected, $C_W$ should be zero with vanish of the downturn of the longitudinal MC at high $T$s.

\begin{figure}
\includegraphics[scale=0.35]{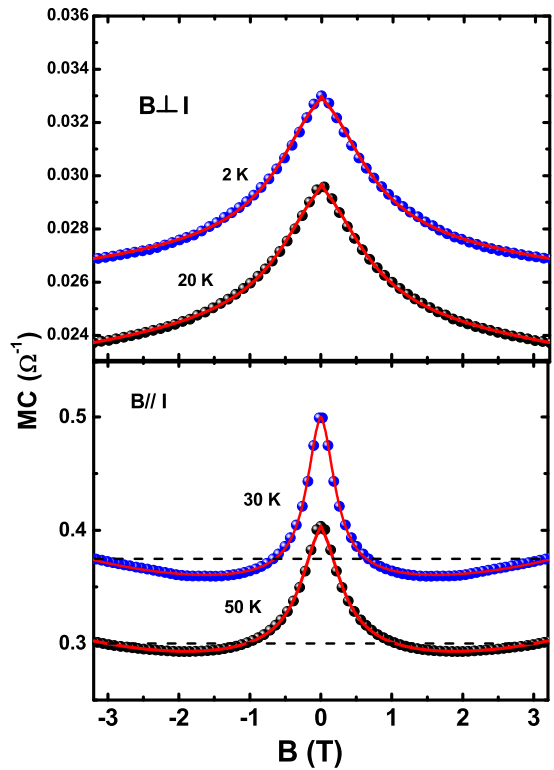}
\caption{\label{fig:fig3}(Color online) (a) Magneto-resistance in the
form of magneto-conductance (MC)
$\delta\sigma_{xx}\equiv\sigma_{x}(B)-\sigma_{xx}(0)\simeq
-MR/\rho^2_{(0)}$ under $P=1.33$ GPa. The red solid lines are the
theoretical fits for different $T$s, see in text. (b) Magneto-conductance MC under $P=1.35$ GPa for selected
$T$s in $B\parallel b$ configuration. The colored solid lines are the
theoretical fits to the Equ. (1) for different $T$s.}
\end{figure}

\textbf{Acknowledgement:} C. Ren wishes to express his thanks to Prof. J. R. Shi and Drs. Jun Zhu and X. C. Huang for valuable discussions. This study is supported by the Ministry of Science and Technology of China (973 project No: 2011CBA00100, 2015CB921303, and 2011CB821404), and Chinese Academy of Sciences (Project ITSNEM) and
the National Science Foundation of China (No.11574373).

$\dag$gfchen@iphy.ac.cn

$\ddag$cong\_ren@iphy.ac.cn

%\end{CJK}
\end{document}